\documentclass[10pt,english]{revtex4}

\usepackage{helvet}
\usepackage{courier}

\usepackage[T1]{fontenc}
\usepackage[latin9]{inputenc}
\usepackage[a4paper]{geometry}
\usepackage{amsmath}

\makeatletter
\@ifundefined{textcolor}{}
{%
 \definecolor{BLACK}{gray}{0}
 \definecolor{WHITE}{gray}{1}
 \definecolor{RED}{rgb}{1,0,0}
 \definecolor{GREEN}{rgb}{0,1,0}
 \definecolor{BLUE}{rgb}{0,0,1}
 \definecolor{CYAN}{cmyk}{1,0,0,0}
 \definecolor{MAGENTA}{cmyk}{0,1,0,0}
 \definecolor{YELLOW}{cmyk}{0,0,1,0}
 }

\makeatother

\usepackage{babel}

\begin{document}

\title{The sinusoid and the phasor}

\author{Kushal Shah}

\email{kushal.shah@weizmann.ac.il}

\affiliation{Faculty of Mathematics and Computer Science, Weizmann Institute of
Science, POB 26, Rehovot 76100, Israel.}

\author{Harishankar Ramachandran}

\email{hsr@ee.iitm.ac.in}

\affiliation{Department of Electrical Engineering, Indian Institute of Technology
Madras, Chennai 600036, India.}
\begin{abstract}
Mathieu equation is widely used to study several natural phenomenon.
In this paper, we show that replacing the sinusoid in the Mathieu
equation with a phasor can lead to solutions that behave in a totally
different way. Solutions of Mathieu equation are either bounded or
grow unboundedly at an exponential rate. Solutions of this new equation
are always unbounded and grow linearly with time.
\end{abstract}
\maketitle
Linear differential equations with periodic coefficients have applications
in many areas like Plasma Physics \cite{Paul,KSHSR} and Condensed
Matter Physics \cite{Kittel}. The general form for such an equation
is\begin{equation}
\dot{\mathbf{x}}=\mathbf{A}(t)\mathbf{x}\label{eq:ODE_general_form}\end{equation}
where $\mathbf{x}$ is a vector and $\mathbf{A}(t)$ is an $n\times n$
matrix with time-periodic coefficients. For $n=2$, a very important
example of such a system of equations is the Mathieu equation \cite{McLachlan,Abram},
\begin{equation}
\ddot{x}=2qx\cos\omega t\label{eq:Mathieu}\end{equation}
which can be written in the matrix form as \begin{equation}
\frac{d}{dt}\left(\begin{array}{c}
x_{1}\\
x_{2}\end{array}\right)=\left(\begin{array}{cc}
0 & 1\\
2q\cos\omega t & 0\end{array}\right)\left(\begin{array}{c}
x_{1}\\
x_{2}\end{array}\right)\label{eq:Mathieu_Equation}\end{equation}
where $q$ is an arbitrary constant and $\omega$ is the oscillation
frequency of the field. 

In many problems of interest in physics and engineering, the sinusoid
is replaced by a phasor since it often makes the analysis easier.
If we replace the sinusoid in Eq. \eqref{eq:Mathieu} by a phasor,
we get \begin{equation}
\frac{d^{2}y}{dt^{2}}=2qye^{j\omega t}\label{eq:eqn_phasor}\end{equation}
where $y$ is a complex function of a real argument, $t$. It must
be noted that Eq. \eqref{eq:eqn_phasor} can also be written in form
of Eq. \eqref{eq:ODE_general_form} with $n=4$. Transforming to the
variable, $\tau=\sqrt{\frac{8q}{\omega^{2}}}e^{jwt/2}$, Eq. \eqref{eq:eqn_phasor}
becomes\begin{equation}
\tau\frac{d^{2}y}{d\tau^{2}}+\frac{dy}{d\tau}+\tau y=0\label{eq:eqn_y_tau}\end{equation}
which is the well known Bessel equation \cite{Abram}. The transformation,
$\tau=\sqrt{\frac{8q}{\omega^{2}}}e^{jwt/2}$, used to get Eq. \eqref{eq:eqn_y_tau}
is not one-to-one, but as we will see later, the solutions have a
branch-cut, which permits the use of this transformation leading to
non-periodic solutions. Equation \eqref{eq:eqn_y_tau} has two linearly
independent solutions and any general solution can be written as a
linear combination of these, \begin{equation}
y\left(\tau\right)=a\phi\left(\tau\right)+b\psi\left(\tau\right)\label{eq:general_solution}\end{equation}
where $\phi,\psi$ are Bessel functions,\begin{equation}
\phi\left(\tau\right)=J_{0}\left(\sqrt{\frac{8q}{\omega^{2}}}e^{jwt/2}\right)\qquad\textrm{and}\qquad\psi\left(\tau\right)=Y_{0}\left(\sqrt{\frac{8q}{\omega^{2}}}e^{jwt/2}\right)\label{eq:phi_psi}\end{equation}
The velocity of the particle at any instant of time is given by,\begin{equation}
v(t)=a\phi'(t)+b\psi'(t)\label{eq:vel_soln}\end{equation}
where, $'$ represents differentiation w.r.t. $t$ . 

In order to find the path of the particles in phase space, we need
to evaluate the bessel functions and their derivatives. The derivative
of $J_{0}$ is $-J_{1}$ and that of $Y_{0}$ is $-Y_{1}$. Though
all these functions are analytic for complex arguments, $Y_{0}$ and
$Y_{1}$ have a branch cut. $Y_{0},Y_{1}$ can be evaluated using
the expression \cite{Gray} \begin{equation}
Y_{0}\left(\tau\right)=\frac{2}{\pi}\left[\left(C+\log\frac{\tau}{2}\right)J_{0}\left(\tau\right)+2\sum_{n=1}^{N}(-1)^{n-1}\frac{J_{2n}\left(\tau\right)}{n}\right]\label{eq:Y_0}\end{equation}
and \begin{equation}
Y_{1}\left(\tau\right)=\frac{2}{\pi}\left[\left(C+\log\frac{\tau}{2}\right)J_{1}\left(\tau\right)-\frac{1}{z}J_{0}\left(\tau\right)+\sum_{n=1}^{N}\frac{(-1)^{n}}{n}\left(J_{2n-1}\left(\tau\right)-J_{2n+1}\left(\tau\right)\right)\right]\label{eq:Y_1}\end{equation}
where, $C=0.577215665$ is the Euler's constant . Although $Y_{0}\left(\tau\right)$
and $Y_{1}\left(\tau\right)$ seem to be a periodic functions of $t$,
the condition that $y$ and $v$ must be continuous and differentiable
imposes certain restrictions. We need to take special care of the
branch cut in $\log\frac{\tau}{2}$ term. In our problem, \begin{eqnarray}
\tau & = & \sqrt{\frac{8q}{\omega^{2}}}e^{jwt/2}\label{eq:z}\end{eqnarray}
which implies \begin{eqnarray}
\log\frac{\tau}{2} & = & \log\sqrt{\frac{4q}{\omega^{2}}}+\frac{jwt}{2}\label{eq:logz2}\end{eqnarray}
From Eq. \eqref{eq:logz2}, it can be seen that the continuity and
differentiability condition forces $\psi(t)$ to be growing in time.
Equation \eqref{eq:logz2} predicts this unbounded growth with time
to be linear, whereas unbounded solutions of the Mathieu equation,
Eq. \eqref{eq:Mathieu}, would have exponential-in-time growth. Also,
Eq. \eqref{eq:eqn_phasor} has no bounded solutions for any value
of $q$, whereas all solutions of Eq. \eqref{eq:Mathieu} are bounded
if $\left|q\right|$ is below a certain critical threshold.

One of the authors (KS) would like to acknowledge support of the Israel
Science Foundation and the Minerva Foundation.

\end{document}